\documentclass[fleqn,usenatbib]{mnras}
\usepackage{newtxtext,newtxmath}
\usepackage[T1]{fontenc}
\DeclareRobustCommand{\VAN}[3]{#2}
\let\VANthebibliography\thebibliography
\def\thebibliography{\DeclareRobustCommand{\VAN}[3]{##3}\VANthebibliography}
\usepackage{graphicx}	
\usepackage{amsmath}	

\def\modify#1{#1}

\title[Revisiting the QMR]{Revisiting the quasi-molecular mechanism of recombination}

\author[Zhiqi Huang]{
Zhiqi Huang$^{1,2}$\thanks{E-mail: huangzhq25@mail.sysu.edu.cn}
\\
${}^{1}$ School of Physics and Astronomy, Sun Yat-sen University, 2 Daxue Road, Tangjia, Zhuhai, 519082, China \\
${}^{2}$ CSST Science Center for the Guangdong-Hongkong-Macau Greater Bay Area, Sun Yat-sen University, Zhuhai, 519082, China 
}

\date{Accepted XXX. Received YYY; in original form ZZZ}

\pubyear{2022}

\begin{document}
\label{firstpage}
\pagerange{\pageref{firstpage}--\pageref{lastpage}}
\maketitle

\begin{abstract}
  The quasi-molecular mechanism of recombination, recently suggested by Kereselidze et al., is a non-standard process where an electron and two neighboring protons in the early universe directly form an ionized hydrogen molecule in a highly excited state, which then descends to lower levels or dissociates. It has been suggested that the increased binding energy due to the participation of a second proton may lead to an earlier cosmic recombination that alleviates the Hubble tension. Revisiting the quasi-molecular channel of recombination in more details, we find that the original work significantly overestimated the probability of finding a pair of adjacent protons in the relevant epoch ($z\sim $ a few thousand). Our new estimation suggests that the quasi-molecular mechanism of recombination cannot be the primary cause of the Hubble tension.  
\end{abstract}



\section{Introduction}

In the standard $\Lambda$ cold dark matter ($\Lambda$CDM) model, the cosmic microwave background (CMB) data implies a Hubble constant value that is  about $5\%$ lower than that from the local distance-ladder measurements. The discordance between the two measurements, commonly dubbed as the Hubble tension, has persisted for almost a decade  and is approaching $\sim 5\sigma$ recently~\citep{Planck2018Params, Riess21, Riess20}. The Hubble tension has inspired many theoretical speculations, such as early dark energy~\citep[e.g.][]{Karwal_2016, Poulin_2019}, modified gravity~\citep[e.g.][]{Sola_2019, Raveri_2020, Yan_2020, Sola_2020}, primordial magnetic fields~\citep{Jedamzik_2020}, and non-standard recombination~\citep{Chiang18, Liu19, Ye20, Sekiguchi20}. See~\citet{Verde19, Knox20, H0Tension_Review, Shah21, Cai22, Cai21} for recent reviews.

The cosmic recombination, especially the hydrogen recombination at redshift $z\sim 1100$, plays a key role in the theoretical calculation of the statistics of the anisotropy of the CMB. The epoch of hydrogen recombination determines the sound horizon at photon last scattering, which combined with the observed acoustic oscillations in the CMB power spectrum gives a precise measurement of the distance to the last scattering surface, and hence the Hubble constant. Moreover, the abundance of free electrons determines the diffusion scale of photons, imprinted on the damping tail of the primordial component of the CMB power spectrum~\citep{Silk68}, which in principle also affects the constraint on the Hubble constant via the degeneracy between cosmological parameters. 

According to the pioneer works by Peebles et al.~\citep{Peebles68, Zeldovich69}, the hydrogen recombination is dominated by the $2s\rightarrow 1s$ two-photon radiative process and cosmological redshift of the Lyman-$\alpha$ line. Till today the basic picture remains unchanged. Peebles' ``effective three-level atom'' approximation, which treats hydrogen as an atom with ground state, first excited state (energy level $n=2$), and continuum (collection of $n>2$ states) that is assumed to be in equilibrium with the radiation, turns out to be a good approximation and was the standard methodology in almost thirty years. In the late 90s, Seager, Sasselov and Scott conducted the first modern calculation beyond Peebles' three-level approximation. By evolving hundreds of atomic energy level populations without assuming their equilibrium with the radiation, \citet{Seager00} improved the accuracy of recombination calculation to a percent level. The corresponding code RECFAST, which uses a fudge factor $F=1.14$ to account for the non-equilibrium between the radiation and the hydrogen $n>2$ levels~\citep{Seager99}, is still used today in many scenarios. The current state-of-the-art codes COSMOREC~\citep{CosmoREC, COSMOSPEC} and HYREC~\citep{HYREC, HYREC-2} further include a variety of detailed physics that matter at a sub-percent level of accuracy. See~\citet{HYREC-2} for a more detailed review.

The quasi-molecular mechanism of recombination (QMR) is yet another recently proposed refinement of the standard picture of recombination~\citep{Kereselidze19}. In the QMR picture, two adjacent protons in the early Universe capture a free electron and directly form a $\mathrm{H_2^+}$ ion at a highly excited state, which then cascades to the ground state of $\mathrm{H_2^+}$ or dissociates to an excited hydrogen atom and a proton. The complexity of the $\mathrm{H_2^+}$ ion makes it difficult to calculate QMR with a full kinematic approach. While significant progress has been made~\citep{Kereselidze20, Kereselidze21, Kereselidze22}, a complete calculation scheme is not yet available, and thus QMR has not been included in any recombination code in the market.

The participation of a second proton increases the binding energy, which may lead to an earlier recombination (smaller sound horizon) and hence a tempting possibility of resolving the Hubble tension within the standard $\Lambda$CDM model~\citep{Beradze21}. However, because the QMR process is a three-body reaction whose phase-space factor is suppressed, naively one would expect its contribution to the total recombination of hydrogen to be too small to account for the $5\%$ discrepancy in the early and late measurements of the Hubble constant. On the contrary, a series of work have argued, based on a very crude estimation, that a substantial QMR contribution is possible~\citep{Kereselidze19, Kereselidze20, Beradze21}. The purpose of the present work is then to revisit the estimation of the QMR contribution with a more detailed calculation.

\section{Average distance between free protons \label{sec:dis}}

Following~\citet{Kereselidze19}, we begin by estimating the average distance between free protons, $\bar{R} = n_p^{-1/3}$, where $n_p$ is the physical number density of free protons. We will firstly assume that the QMR contribution is negligible. The validity of this assumption is checked in the next section.

At redshift $z \lesssim 10^4$, the comoving number density of hydrogen nuclei (free or bounded in hydrogen atoms) is conserved. Thus the physical number density of free protons can be written as
\begin{equation}
  n_p = n_0(1+z)^3x_p, \label{eq:np}
\end{equation}
where $x_p$ is the ionization fraction of hydrogen, $n_0$ is the comoving number density of hydrogen nuclei, given by
\begin{equation}
  n_0 = \frac{3\Omega_bH_0^2(1-Y_P)}{8\pi G m_H}.\label{eq:n0}
\end{equation}
Substituting the Planck best-fit~\citep{Planck2018Params} baryon fraction $\Omega_b = 0.04930$, Hubble constant $H_0 = 67.36\mathrm{km/s/Mpc}$, helium mass fraction $Y_P=0.2454$, as well as the Newton's constant $G$ and the hydrogen mass $m_H$ into Eq.~\eqref{eq:n0}, we obtain  $n_0 \approx 0.2\mathrm{m^{-3}}$. Thus, the average distance between free protons at redshift $z$ is at least of order $\sim \frac{1}{1+z}$ meters, which for $z\lesssim 10^4$ is $\sim$ million times larger than the Bohr radius $a_0\approx 5.3\times 10^{-11}\mathrm{m}$.

\begin{figure}
  \includegraphics[width=\columnwidth]{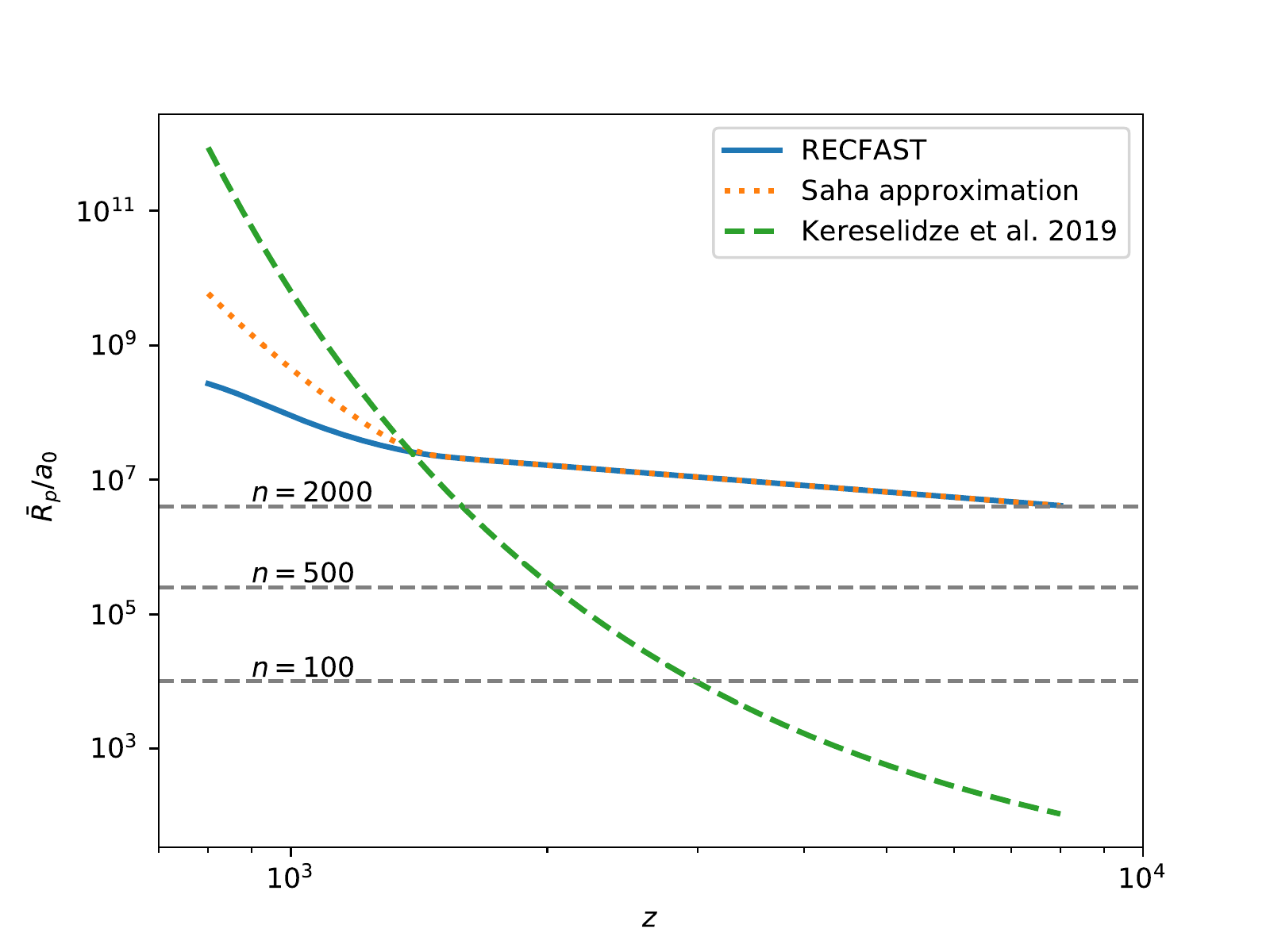}
  \caption{Average distance between free protons as a function of redshift. The solid bue line is calculated with a modified RECFAST code. The orange dotted line is calculated with Saha approximation. The dashed green line is a reproduction of the red line in Figure~1 of \citet{Kereselidze19}, which is calculated from Eq.~(5) of \citet{Kereselidze19} with the assumption $x_e\approx x_H$. \modify{The horizontal gray dashed lines show the radius of the $n$-th shell of the hydrogen atom ($\frac{r_n}{a_0} = n^2$).} \label{fig:Rp}}
\end{figure}

The RECFAST code can be easily modified to output the $n_p(z)$ function, and hence the ratio $\frac{\bar{R}_p}{a_0}$. The numerical result shown as the solid blue line in Figure~\ref{fig:Rp} confirms our previous statement $\bar{R}_p\gtrsim 10^6 a_0$. 

Although the numerical RECFAST output is accurate, it does not elucidate what has led to the incorrect estimation of $\bar{R}_p$ in \citet{Kereselidze19}, where Saha approximation was used.
For a line-to-line comparison, we now apply Saha approximation to re-estimate $\bar{R}_p$.

At thermal equilibrium  with temperature $T$, the number density of a particle species $a$ with mass $m_a$, chemical potential $\mu_a$, and intrinsic degrees of freedom $g_a$ is given by
\begin{equation}
  n_a = g_a \left(\frac{m_a k_BT}{2\pi \hbar^2}\right)^{3/2}e^{-\frac{m_ac^2-\mu_a}{k_BT}}. \label{eq:n}
\end{equation}
For proton, the conservation of hydrogen nuclei plays a key role in the determination of the chemical potential $\mu_p$, which in the cosmological context is comparable to its mass energy $m_pc^2$. The chemical potential was incorrectly ignored in Eq.~(1) of \citet{Kereselidze19}, making a wrong impression that the number density of free protons is mainly determined by the background temperature.

For the hydrogen recombination reaction
\begin{equation}
  p + e \leftrightarrow \mathrm{H} + \gamma, \label{eq:Hrec}
\end{equation}
the detailed balance $\mu_p+\mu_e = \mu_H$ leads to the Saha approximation
\begin{equation}
  \frac{n_pn_e}{n_H} = \left(\frac{m_ek_BT}{2\pi\hbar^2}\right)^{3/2}e^{-\frac{E_H}{k_BT}}, \label{eq:Saha_H}
\end{equation}
where $E_H = m_p+m_e-m_{\rm H} = 13.6\,\mathrm{eV}$ is the hydrogen ionization energy. Here $n_e$, $n_p$ and $n_H$ are physical number densities of free protons, free electrons, and ground-state hydrogen atoms, respectively.

To estimate the order of magnitude of $n_p$ from Eq.~\eqref{eq:Saha_H}, \citet{Kereselidze19} assumed $\frac{n_e}{n_H}\sim O(1)$, which is actually incorrect for most of the redshift range. In particular at $z>2000$, most hydrogen atoms are ionized and therefore $\frac{n_e}{n_H} \gg 1$. Incorrectly taking $\frac{n_e}{n_H}\sim O(1)$ in \eqref{eq:Saha_H} leads to an overestimated $n_p$ and hence an underestimated $\bar{R}_p$ \modify{at $z> 2000$}.

\modify{Although the chemical potentials cancel in the combination $\frac{n_pn_e}{n_H}$ in Saha equation~\eqref{eq:Saha_H}, the number density of a single species (e.g., $n_p$) still relies on its chemical potential, which physically is determined by the conservation of hydrogen nuclei in a comoving volume. Thus,} the standard approach to calculate hydrogen recombination with Saha approximation is to combine the conservation of hydrogen nuclei $n_e+n_H\approx n_0(1+z)^3$  and the obvious equality $n_e\approx n_p$ with Eq.~\eqref{eq:Saha_H}. For a comparison, we also plot $\bar{R}_p(z)$ from the standard Saha approach in Fig~\ref{fig:Rp}.

Our result indicates that the average distance between free protons is much greater than the Bohr radius during cosmological recombination epoch. \modify{For $800<z<8000$, $\bar{R}_p$ is greater than the radius of the 2000-th shell of the hydrogen atom ($\bar{R}_p > r_{2000} = 4\times 10^6 a_0$). It has been shown in \citet{CosmoREC} that  shells with $n\gtrsim 100$ only contribute a sub-percent level correction to the recombination dynamics at $z>800$. Thus, for the purpose of finding the primary cause of the Hubble tension, which is a $\sim 5\%$ effect, we may only focus on proton pairs with the distance between the two protons $\lesssim r_{100}=10^4a_0$. Assuming Poisson statistics, it can be easily estimated that at $z=8000$ less than $3\times 10^{-8}$ fraction of protons has a neighboring proton within a distance $<10^4 a_0$, and at $z=2000$ the fraction drops to $\lesssim 10^{-10}$.}

Given the tiny chance of finding adjacent proton pairs, we are in the position to question whether QMR can have any visible impact on the recombination history at all. In the next section, we proceed to estimate an upper bound of the QMR contribution.

\section{The QMR contribution\label{sec:QMR}}

Figure~\ref{fig:Rp} suggests that the Saha approximation is inaccurate for $z\lesssim 1500$, where the assumption of thermal equilibrium fails. In fact, even the reaction~\eqref{eq:Hrec} (direct formation of a ground-state hydrogen atom) is not the dominant channel of hydrogen recombination~\citep{Peebles68}. The idea of Saha approximation is to consider Eq.~\eqref{eq:Hrec} as an effective reaction connecting the initial and final states. Despite often being inaccurate in cosmological scenarios, Saha approximation typically suffices for the purpose of getting a rough order-of-magnitude estimation.

Because a complete kinetic approach for QMR is not yet available,  we take a step back and use Saha approximation to roughly estimate the QMR contribution. We do not consider the case where the final product is a hydrogen atom, because in Saha-approximation approach it is included in the effective reaction~\eqref{eq:Hrec}. Assuming thermal equilibrium, $\mathrm{H}_2^+$ in exicted states are less abundant than that in the ground-state. Thus, for an order-of-magnitude estimation, we only consider $\mathrm{H}_2^+$ in the ground-state. (The same philosophy has been used in the standard Saha approximation for hydrogen atoms and helium atoms, too.)

With Saha approximation, the production/dissociation of $\mathrm{H}_2^+$ ions can be considered as an effective reaction
\begin{equation}
  e+p+p \leftrightarrow \mathrm{H}_2^+ + \gamma. 
\end{equation}

Applying Eq.~\eqref{eq:n} to $e$, $p$ and $\mathrm{H}_2^+$, and using the detailed balance $\mu_e + 2\mu_p = \mu_{H_2^+}$, we obtain a three-body Saha equation
\begin{equation}
  \frac{n_en_p^2}{n_{H_2^+}} = \left(\frac{m_em_p}{2}\right)^{3/2}\left(\frac{k_BT}{2\pi\hbar^2}\right)^3e^{-\frac{E_{H_2^+}}{k_BT}}, \label{eq:nH2}
\end{equation}
where $E_{H_2^+} = m_e+2m_p - m_{H_2^+}= 16.4\,\mathrm{eV}$ is the total binding energy of $\mathrm{H_2^+}$.

The impact of QMR can be characterized by the abundance of $\mathrm{H_2^+}$, which is defined as the ratio of $n_{H_2^+}$ to the total number density of hydrogen nuclei,
\begin{equation}
  x_{H_2^+}(z) \equiv \frac{n_{H_2^+}(z)}{n_0(1+z)^3}. \label{eq:xH2}
\end{equation}
\modify{The electron and proton abundances $x_e, x_p$ are defined in a similar manner by replacing $H_2^+$ in Eq.~\eqref{eq:xH2} with $e, p$, respectively.}

\begin{figure}
  \includegraphics[width=\columnwidth]{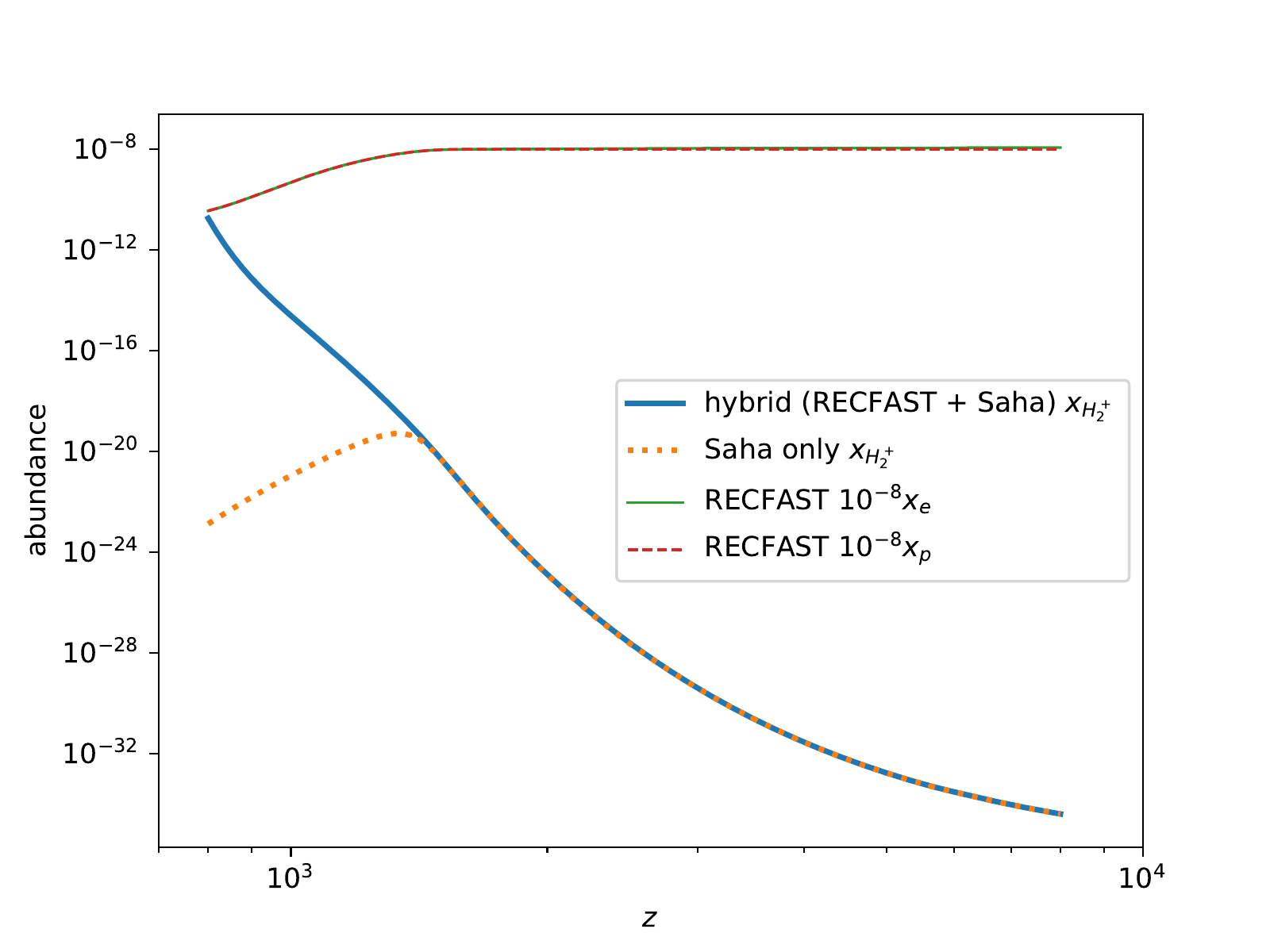}
  \caption{The abundance of $\mathrm{H_2^+}$ as a function of redshift. The thick blue line is calculated with a hybrid approach, where numerical $n_e(z)$ and $n_p(z)$ from RECFAST are used in the three-body Saha equation~\eqref{eq:nH2}. The orange dotted line is a pure Saha approximation calculation, where $n_e(z)$ and $n_p(z)$ in the three-body Saha equation~\eqref{eq:nH2} are computed with two-body Saha equations for hydrogen and helium atoms. The thin green line and dashed red line are electron and proton abundances rescaled by a factor $10^{-8}$.\label{fig:xH2}}
\end{figure}

To calculate $n_{H_2^+}(z)$ from the three-body Saha equation~\eqref{eq:nH2}, we need to know $n_e(z)$ and $n_p(z)$, which can be obtained numerically from RECFAST. \modify{As shown in Figure~\ref{fig:xH2}, for $z>800$, the abundance of $H_2^+$ is much smaller than that of electron (or proton) by a factor of $\lesssim 10^{-8}$. Thus, the correction to RECFAST $n_e(z)$ and $n_p(z)$ due to QMR can be safely ignored, validating our usage of RECFAST $n_e(z)$ and $n_p(z)$ in Eq.~\eqref{eq:nH2}.}

Our hybrid approach (Saha + RECFAST) guarantees the accuracy in $n_e(z)$ and $n_p(z)$, thus is more accurate than a pure Saha estimation that is also shown in Figure~\ref{fig:xH2} for a comparison.

\section{Discussion and Conclusions\label{sec:conclu}}

The previous studies of QMR is largely motivated by the incorrect physical picture that the average distance between free protons $\bar{R}_p$ (at $z\lesssim 10^4$) can be comparable to the Bohr radius. While $\bar{R}_p$ is now correctly estimated to be at least $\sim$ millions times larger than the Bohr radius, it is natural to ask whether the QMR channel can have any significant influence on the recombination history. Using a hybrid approach (Saha approximation + RECFAST numerics) we find the $\mathrm{H}_2^+$ abundance is $\lesssim 10^{-11}$ for the relevant redshift range ($800<z<10^4$), which is far too small to explain the Hubble tension.

Opposite conclusions have been drawn in the present work and in \citet{Beradze21}. The main physics can be understood as follows. The participation of a second proton increases the binding energy, but also suppresses the phase-space factor. Both effects are naturally encoded in the three-body Saha equation used in the present work, while the effective two-body Saha equation used in \citet{Beradze21} fails to capture the suppression of the phase-space factor. See Eq.~\eqref{eq:nH2} in the present work and Eq.~(9) in \citet{Beradze21} for a more detailed comparison.

The $\mathrm{H}_2^+$ ion (and $\mathrm{H}_2$ molecules etc.) can also be formed via two-body channels during the recombination epoch. The contribution of these channels are calculated in details in \citet{Alizadeh11}, where a similar upper bound $\sim 10^{-13}$ has been found for the abundance of hydrogen molecules. Thus, we may conclude that as far as cosmological recombination is concerned, hydrogen molecules can be safely ignored.

\section*{Acknowledgements}

This work is supported by the National key R\&D Program of China (Grant No. 2020YFC2201600), National SKA Program of China No. 2020SKA0110402, National Natural Science Foundation of China (NSFC) under Grant No. 12073088, and Guangdong Major Project of Basic and Applied Basic Research (Grant No. 2019B030302001).

\section*{Data Availability}

The data (source codes) underlying this article are available at \url{http://zhiqihuang.top/codes/qmr.tar.gz}.


\begin{thebibliography}{}
\makeatletter
\relax
\def\mn@urlcharsother{\let\do\@makeother \do\$\do\&\do\#\do\^\do\_\do\%\do\~}
\def\mn@doi{\begingroup\mn@urlcharsother \@ifnextchar [ {\mn@doi@}
  {\mn@doi@[]}}
\def\mn@doi@[#1]#2{\def\@tempa{#1}\ifx\@tempa\@empty \href
  {http://dx.doi.org/#2} {doi:#2}\else \href {http://dx.doi.org/#2} {#1}\fi
  \endgroup}
\def\mn@eprint#1#2{\mn@eprint@#1:#2::\@nil}
\def\mn@eprint@arXiv#1{\href {http://arxiv.org/abs/#1} {{\tt arXiv:#1}}}
\def\mn@eprint@dblp#1{\href {http://dblp.uni-trier.de/rec/bibtex/#1.xml}
  {dblp:#1}}
\def\mn@eprint@#1:#2:#3:#4\@nil{\def\@tempa {#1}\def\@tempb {#2}\def\@tempc
  {#3}\ifx \@tempc \@empty \let \@tempc \@tempb \let \@tempb \@tempa \fi \ifx
  \@tempb \@empty \def\@tempb {arXiv}\fi \@ifundefined
  {mn@eprint@\@tempb}{\@tempb:\@tempc}{\expandafter \expandafter \csname
  mn@eprint@\@tempb\endcsname \expandafter{\@tempc}}}

\bibitem[\protect\citeauthoryear{Aghanim et~al.}{Aghanim
  et~al.}{2020}]{Planck2018Params}
Aghanim N.,  et~al., 2020, \mn@doi [Astron. Astrophys.]
  {10.1051/0004-6361/201833910}, 641, A6

\bibitem[\protect\citeauthoryear{{Ali-Ha{\"\i}moud} \&
  {Hirata}}{{Ali-Ha{\"\i}moud} \& {Hirata}}{2011}]{HYREC}
{Ali-Ha{\"\i}moud} Y.,  {Hirata} C.~M.,  2011, \mn@doi [\prd]
  {10.1103/PhysRevD.83.043513}, \href
  {https://ui.adsabs.harvard.edu/abs/2011PhRvD..83d3513A} {83, 043513}

\bibitem[\protect\citeauthoryear{{Alizadeh} \& {Hirata}}{{Alizadeh} \&
  {Hirata}}{2011}]{Alizadeh11}
{Alizadeh} E.,  {Hirata} C.~M.,  2011, \mn@doi [\prd]
  {10.1103/PhysRevD.84.083011}, \href
  {https://ui.adsabs.harvard.edu/abs/2011PhRvD..84h3011A} {84, 083011}

\bibitem[\protect\citeauthoryear{{Beradze} \& {Gogberashvili}}{{Beradze} \&
  {Gogberashvili}}{2021}]{Beradze21}
{Beradze} R.,  {Gogberashvili} M.,  2021, \mn@doi [Physics of the Dark
  Universe] {10.1016/j.dark.2021.100841}, \href
  {https://ui.adsabs.harvard.edu/abs/2021PDU....3200841B} {32, 100841}

\bibitem[\protect\citeauthoryear{{Cai}, {Guo}, {Wang}, {Yu}  \& {Zhou}}{{Cai}
  et~al.}{2022a}]{Cai22}
{Cai} R.-G.,  {Guo} Z.-K.,  {Wang} S.-J.,  {Yu} W.-W.,   {Zhou} Y.,  2022a,
  arXiv e-prints, \href {https://ui.adsabs.harvard.edu/abs/2022arXiv220212214C}
  {p. arXiv:2202.12214}

\bibitem[\protect\citeauthoryear{{Cai}, {Guo}, {Wang}, {Yu}  \& {Zhou}}{{Cai}
  et~al.}{2022b}]{Cai21}
{Cai} R.-G.,  {Guo} Z.-K.,  {Wang} S.-J.,  {Yu} W.-W.,   {Zhou} Y.,  2022b,
  \mn@doi [\prd] {10.1103/PhysRevD.105.L021301}, \href
  {https://ui.adsabs.harvard.edu/abs/2022PhRvD.105b1301C} {105, L021301}

\bibitem[\protect\citeauthoryear{{Chiang} \& {Slosar}}{{Chiang} \&
  {Slosar}}{2018}]{Chiang18}
{Chiang} C.-T.,  {Slosar} A.,  2018, arXiv e-prints, \href
  {https://ui.adsabs.harvard.edu/abs/2018arXiv181103624C} {p. arXiv:1811.03624}

\bibitem[\protect\citeauthoryear{{Chluba} \& {Ali-Ha{\"\i}moud}}{{Chluba} \&
  {Ali-Ha{\"\i}moud}}{2016}]{COSMOSPEC}
{Chluba} J.,  {Ali-Ha{\"\i}moud} Y.,  2016, \mn@doi [\mnras]
  {10.1093/mnras/stv2691}, \href
  {https://ui.adsabs.harvard.edu/abs/2016MNRAS.456.3494C} {456, 3494}

\bibitem[\protect\citeauthoryear{Chluba \& Thomas}{Chluba \&
  Thomas}{2011}]{CosmoREC}
Chluba J.,  Thomas R.~M.,  2011, \mn@doi [Mon. Not. Roy. Astron. Soc.]
  {10.1111/j.1365-2966.2010.17940.x}, 412, 748

\bibitem[\protect\citeauthoryear{Di~Valentino et~al.,}{Di~Valentino
  et~al.}{2021}]{H0Tension_Review}
Di~Valentino E.,  et~al., 2021, \mn@doi [Class. Quant. Grav.]
  {10.1088/1361-6382/ac086d}, 38, 153001

\bibitem[\protect\citeauthoryear{{Jedamzik} \& {Pogosian}}{{Jedamzik} \&
  {Pogosian}}{2020}]{Jedamzik_2020}
{Jedamzik} K.,  {Pogosian} L.,  2020, \mn@doi [Phys. Rev. Lett.]
  {10.1103/PhysRevLett.125.181302}, \href
  {https://ui.adsabs.harvard.edu/abs/2020PhRvL.125r1302J} {125, 181302}

\bibitem[\protect\citeauthoryear{Karwal \& Kamionkowski}{Karwal \&
  Kamionkowski}{2016}]{Karwal_2016}
Karwal T.,  Kamionkowski M.,  2016, \mn@doi [Physical Review D]
  {10.1103/physrevd.94.103523}, 94

\bibitem[\protect\citeauthoryear{{Kereselidze} \& {Noselidze}}{{Kereselidze} \&
  {Noselidze}}{2021}]{Kereselidze22}
{Kereselidze} T.,  {Noselidze} I.,  2021, arXiv e-prints, \href
  {https://ui.adsabs.harvard.edu/abs/2021arXiv210411584K} {p. arXiv:2104.11584}

\bibitem[\protect\citeauthoryear{{Kereselidze}, {Noselidze}  \&
  {Ogilvie}}{{Kereselidze} et~al.}{2019}]{Kereselidze19}
{Kereselidze} T.,  {Noselidze} I.,   {Ogilvie} J.~F.,  2019, \mn@doi [\mnras]
  {10.1093/mnras/stz1808}, \href
  {https://ui.adsabs.harvard.edu/abs/2019MNRAS.488.2093K} {488, 2093}

\bibitem[\protect\citeauthoryear{{Kereselidze}, {Noselidze}  \&
  {Ogilvie}}{{Kereselidze} et~al.}{2021}]{Kereselidze20}
{Kereselidze} T.,  {Noselidze} I.,   {Ogilvie} J.~F.,  2021, \mn@doi [\mnras]
  {10.1093/mnras/staa3622}, \href
  {https://ui.adsabs.harvard.edu/abs/2021MNRAS.501.1160K} {501, 1160}

\bibitem[\protect\citeauthoryear{{Kereselidze}, {Noselidze}  \&
  {Ogilvie}}{{Kereselidze} et~al.}{2022}]{Kereselidze21}
{Kereselidze} T.,  {Noselidze} I.,   {Ogilvie} J.~F.,  2022, \mn@doi [\mnras]
  {10.1093/mnras/stab3102}, \href
  {https://ui.adsabs.harvard.edu/abs/2022MNRAS.509.1755K} {509, 1755}

\bibitem[\protect\citeauthoryear{{Knox} \& {Millea}}{{Knox} \&
  {Millea}}{2020}]{Knox20}
{Knox} L.,  {Millea} M.,  2020, \mn@doi [\prd] {10.1103/PhysRevD.101.043533},
  \href {https://ui.adsabs.harvard.edu/abs/2020PhRvD.101d3533K} {101, 043533}

\bibitem[\protect\citeauthoryear{{Lee} \& {Ali-Ha{\"\i}moud}}{{Lee} \&
  {Ali-Ha{\"\i}moud}}{2020}]{HYREC-2}
{Lee} N.,  {Ali-Ha{\"\i}moud} Y.,  2020, \mn@doi [\prd]
  {10.1103/PhysRevD.102.083517}, \href
  {https://ui.adsabs.harvard.edu/abs/2020PhRvD.102h3517L} {102, 083517}

\bibitem[\protect\citeauthoryear{Liu, Huang, Luo, Miao, Singh  \& Huang}{Liu
  et~al.}{2020}]{Liu19}
Liu M.,  Huang Z.,  Luo X.,  Miao H.,  Singh N.~K.,   Huang L.,  2020, \mn@doi
  [Sci. China Phys. Mech. Astron.] {10.1007/s11433-019-1509-5}, 63, 290405

\bibitem[\protect\citeauthoryear{{Peebles}}{{Peebles}}{1968}]{Peebles68}
{Peebles} P.~J.~E.,  1968, \mn@doi [\apj] {10.1086/149628}, \href
  {https://ui.adsabs.harvard.edu/abs/1968ApJ...153....1P} {153, 1}

\bibitem[\protect\citeauthoryear{Poulin, Smith, Karwal  \& Kamionkowski}{Poulin
  et~al.}{2019}]{Poulin_2019}
Poulin V.,  Smith T.~L.,  Karwal T.,   Kamionkowski M.,  2019, \mn@doi
  [Physical Review Letters] {10.1103/physrevlett.122.221301}, 122

\bibitem[\protect\citeauthoryear{{Raveri}}{{Raveri}}{2020}]{Raveri_2020}
{Raveri} M.,  2020, \mn@doi [\prd] {10.1103/PhysRevD.101.083524}, \href
  {https://ui.adsabs.harvard.edu/abs/2020PhRvD.101h3524R} {101, 083524}

\bibitem[\protect\citeauthoryear{{Riess} et~al.,}{{Riess}
  et~al.}{2021a}]{Riess21}
{Riess} A.~G.,  et~al., 2021a, arXiv e-prints, \href
  {https://ui.adsabs.harvard.edu/abs/2021arXiv211204510R} {p. arXiv:2112.04510}

\bibitem[\protect\citeauthoryear{Riess, Casertano, Yuan, Bowers, Macri, Zinn
  \& Scolnic}{Riess et~al.}{2021b}]{Riess20}
Riess A.~G.,  Casertano S.,  Yuan W.,  Bowers J.~B.,  Macri L.,  Zinn J.~C.,
  Scolnic D.,  2021b, \mn@doi [Astrophys. J. Lett.] {10.3847/2041-8213/abdbaf},
  908, L6

\bibitem[\protect\citeauthoryear{{Seager}, {Sasselov}  \& {Scott}}{{Seager}
  et~al.}{1999}]{Seager99}
{Seager} S.,  {Sasselov} D.~D.,   {Scott} D.,  1999, \mn@doi [\apjl]
  {10.1086/312250}, \href
  {https://ui.adsabs.harvard.edu/abs/1999ApJ...523L...1S} {523, L1}

\bibitem[\protect\citeauthoryear{{Seager}, {Sasselov}  \& {Scott}}{{Seager}
  et~al.}{2000}]{Seager00}
{Seager} S.,  {Sasselov} D.~D.,   {Scott} D.,  2000, \mn@doi [\apjs]
  {10.1086/313388}, \href
  {https://ui.adsabs.harvard.edu/abs/2000ApJS..128..407S} {128, 407}

\bibitem[\protect\citeauthoryear{Sekiguchi \& Takahashi}{Sekiguchi \&
  Takahashi}{2021}]{Sekiguchi20}
Sekiguchi T.,  Takahashi T.,  2021, \mn@doi [Phys. Rev. D]
  {10.1103/PhysRevD.103.083507}, 103, 083507

\bibitem[\protect\citeauthoryear{{Shah}, {Lemos}  \& {Lahav}}{{Shah}
  et~al.}{2021}]{Shah21}
{Shah} P.,  {Lemos} P.,   {Lahav} O.,  2021, \mn@doi [\aapr]
  {10.1007/s00159-021-00137-4}, \href
  {https://ui.adsabs.harvard.edu/abs/2021A&ARv..29....9S} {29, 9}

\bibitem[\protect\citeauthoryear{{Silk}}{{Silk}}{1968}]{Silk68}
{Silk} J.,  1968, \mn@doi [\apj] {10.1086/149449}, \href
  {https://ui.adsabs.harvard.edu/abs/1968ApJ...151..459S} {151, 459}

\bibitem[\protect\citeauthoryear{{Sol{\`a} Peracaula}, {G{\'o}mez-Valent}, {de
  Cruz P{\'e}rez}  \& {Moreno-Pulido}}{{Sol{\`a} Peracaula}
  et~al.}{2019}]{Sola_2019}
{Sol{\`a} Peracaula} J.,  {G{\'o}mez-Valent} A.,  {de Cruz P{\'e}rez} J.,
  {Moreno-Pulido} C.,  2019, \mn@doi [\apjl] {10.3847/2041-8213/ab53e9}, \href
  {https://ui.adsabs.harvard.edu/abs/2019ApJ...886L...6S} {886, L6}

\bibitem[\protect\citeauthoryear{Sol\`a~Peracaula, G\'omez-Valent, de
  Cruz~P\'erez  \& Moreno-Pulido}{Sol\`a~Peracaula et~al.}{2020}]{Sola_2020}
Sol\`a~Peracaula J.,  G\'omez-Valent A.,  de Cruz~P\'erez J.,   Moreno-Pulido
  C.,  2020, \mn@doi [Class. Quant. Grav.] {10.1088/1361-6382/abbc43}, 37,
  245003

\bibitem[\protect\citeauthoryear{{Verde}, {Treu}  \& {Riess}}{{Verde}
  et~al.}{2019}]{Verde19}
{Verde} L.,  {Treu} T.,   {Riess} A.~G.,  2019, \mn@doi [Nature Astronomy]
  {10.1038/s41550-019-0902-0}, \href
  {https://ui.adsabs.harvard.edu/abs/2019NatAs...3..891V} {3, 891}

\bibitem[\protect\citeauthoryear{Yan, Zhang, Chen, Zhang, Cai  \&
  Saridakis}{Yan et~al.}{2020}]{Yan_2020}
Yan S.-F.,  Zhang P.,  Chen J.-W.,  Zhang X.-Z.,  Cai Y.-F.,   Saridakis E.~N.,
   2020, \mn@doi [Phys. Rev. D] {10.1103/PhysRevD.101.121301}, 101, 121301

\bibitem[\protect\citeauthoryear{Ye \& Piao}{Ye \& Piao}{2020}]{Ye20}
Ye G.,  Piao Y.-S.,  2020, \mn@doi [Phys. Rev. D]
  {10.1103/PhysRevD.101.083507}, 101, 083507

\bibitem[\protect\citeauthoryear{{Zel'dovich}, {Kurt}  \&
  {Syunyaev}}{{Zel'dovich} et~al.}{1969}]{Zeldovich69}
{Zel'dovich} Y.~B.,  {Kurt} V.~G.,   {Syunyaev} R.~A.,  1969, Soviet Journal of
  Experimental and Theoretical Physics, \href
  {https://ui.adsabs.harvard.edu/abs/1969JETP...28..146Z} {28, 146}

\makeatother
\end{thebibliography}

\bsp	
\label{lastpage}
\end{document}